\documentclass[pra,preprint,aps]{revtex4}
\begin{document}
\title{On the Absence of Spurious Eigenstates in an Iterative Algorithm Proposed By Waxman}
\author{R. A. Andrew, H. G. Miller\footnote{E-Mail: hmiller@maple.up.ac.za}, and A. R. Plastino}
\affiliation{Department of Physics, University of Pretoria, Pretoria 0002, South Africa}

\begin{abstract}
 We discuss a remarkable property of an iterative algorithm
for eigenvalue problems recently advanced by Waxman that  
constitutes a clear advantage over other iterative procedures. 
In quantum mechanics, as well as in other fields, it is often 
necessary to deal with operators exhibiting both a continuum 
and a discrete spectrum. For this kind of operators, the problem of 
identifying  spurious eigenpairs which appear in iterative algorithms 
like the Lanczos algorithm does not occur in the algorithm proposed by 
Waxman.

\noindent
PACS\ 03.65.Ge,\ 02.60.Lj 
\end{abstract}
\maketitle
The Hamiltonian operator which describes a quantum mechanical system generally 
possesses both a continuum as well as a discrete spectrum. 
A similar situation also occurs in other fields, such as theoretical 
population genetics \cite{W02}.
In many cases one is only interested in a few of the lower-lying bound 
states of the system. When only bound states are present iterative 
algorithms such as the Lanczos algorithm \cite{L50} yield good approximations to the lower-lying 
eignstates with good convergence properties \cite{SN69,H75,W72,PM95}. 
On the other hand, the presence of the continuum leads to complications which can be 
circumvented but not without introducing spurious eigensolutions 
that need to be identified and eliminated \cite{AM03}. Such spurious 
eigensolutions, however, do not occur in an algorihthm 
recently proposed by Waxman\cite{W98}. We compare the two algorithms 
and demonstrate this behaviour in a simple numerical example.    

The 
presence of the continuum leads to complications  in the Lanczos algorithm\cite{L50}.
  Finding  a suitable 
start  vector  is by no means trivial\cite{KMB81} since the Lanczos algorithm  
can only be applied to states which are normalizable in the  $\mathcal{L}^2$ 
sense. For those  operators which possess  a continuum as well as a point 
spectrum, the space spanned  by the bound state eigenfunctions is by itself 
certainly not complete and  a suitable  start vector should  be composed 
only of components in the subspace spanned by the bound state eigenvectors. 
  Usually the start vector is chosen from a 
complete set of analytic $\mathcal{L}^2$ functions which define a space 
$\mathcal{F}$. This space is in most cases not necessarily of the same dimension 
as the subspace spanned by the exact eigenvectors. On the other hand, if the 
Lanczos algorithm is applied with this  choice  for the start vector, the 
eigenpairs obtained will correspond to those of the operator $\hat{H}$ projected 
onto $\mathcal{F}$. A subset of these eigenstates must correspond to the exact 
eigenpairs of the unprojected Hamiltonian operator since the exact eigenstates 
can be expanded in terms of the complete set of states  which span  
$\mathcal{F}$.  The exact bound states can be identified and separated from the 
spurious bound states in the following 
manner\cite{AM03}. After  each iteration,  for each of the converging eigenpairs 
($e_{l \beta}$,\ $|e_{l \beta}\rangle$), $\Delta_{l \beta}=|e_{l 
\beta}^2-<e_{l \beta}|\hat{H}^2|e_{l \beta}>|$ (where $l$ is the iteration 
number) is calculated and a determination is made  as to whether  $\Delta$ is 
converging  toward zero or not.  For the exact bound states of $\hat{H}$,  
$\Delta$ must be identically zero while the  other spurious eigenstates states of the 
projected operator should converge  to some non-zero positive   value. Provided 
sufficient iterations are performed, it is possible, in this manner to 
identify uniquely the approximate eigenpairs which ultimately will converge to 
the exact bound states. 

In the present note we wish to point out that this difficulty is avoided in a 
recently proposed iterative algorithm for determining the bound state 
eigenpairs of  linear differential operators such as the Schr\"odinger 
Hamiltonian\cite{W98}.  This  algorithm has many  advantages not the  least of 
which is its simplicity and an excellent convergence rate.  The eigenpairs are 
determined as functions of the strength of the potential in the following 
manner.  For simplicity consider a one-dimensional eigenvalue equation\cite{W98}
\begin{equation}
[-\partial^2_{x} -\lambda V(x)] u(x)=-\epsilon u(x) \label{heq}
\end{equation}
\begin{equation}
\lim_{|x|->\infty}u(x)=0
\end{equation}
where $\partial_{x}=\frac{\partial}{\partial_x}$;   $\lambda > 0$ is the 
strength parameter of the attractive potential ($\lambda$V(x) $<$ 0 and 
V(x)$\rightarrow 0 \ as \ |x| \rightarrow  \infty$) and the energy eigenvalue, 
$-\epsilon $ (with $\epsilon  >  0$), is negative and corresponds to a bound 
state. Using Green's method a solution to eq(\ref{heq}) is given by
\begin{equation}
u(x)= \lambda\int^\infty_{-\infty}G_\epsilon(x-x') V(x')u(x')dx' \label{G} 
\end{equation}
where the  Green's function $G_\epsilon(x)$ satisfies
\begin{equation}
[-\partial^2_{x} + \epsilon]G_\epsilon(x)=\delta(x)
\end{equation}
\begin{equation}
\lim_{|x|->\infty}G_\epsilon(x)=0.
\end{equation}
Normalizing u(x) at an arbitrary $x_{ref}$ 
\begin{equation}
u(x_{ref})=1
\end{equation}
allows $\lambda$ to be written as (see eq(\ref{G}))
\begin{equation}
\lambda=\frac{1}{\int G_{\epsilon}(x')v(x')u(x') dx'} \label{l}
\end{equation}
which can then be used to eliminate $\lambda$ from eq(\ref{G})
\begin{equation}
u(x)=\frac{\int^\infty_{-\infty}G_\epsilon(x-x') V(x')u(x')dx'}{\int 
G_{\epsilon}(x')v(x')u(x') dx'}\label{u}.
\end{equation}
Using equations (\ref{l}) and (\ref{u}), $\lambda$ can be determined as a function 
of $\epsilon$ in the following manner. For a particular choice of $\epsilon$ 
eq(\ref{u}) can be 
iterated
\begin{equation}
u_{n+1}(x)=\frac{\int^\infty_{-\infty}G_\epsilon(x-x') 
V(x')u_n(x')dx'}{\int G_{\epsilon}(x')v(x')u_n(x') dx'} 
\label{iu}
\end{equation}
until it converges and $\lambda$ can then be determined 
from eq(\ref{l}). Repeating for different values of $\epsilon$ yields a set of 
different values of the potential strength $\lambda$. When enough points have 
been determined, a simple interpolation procedure yields the dependence of $\epsilon$ on $\lambda$. 
Note that no diagonalization is required. In spite of the necessity of interpolating, the rapid  
convergence of the numerical solution of eq(\ref{iu})  makes the present algorithm extremely viable. 
Furthermore, a proof of coverconvergence has been given and the algorithm can be extended for the calculation 
of excited states\cite{W98}.

In order to demonstrate that spurious solutions do not  occur in the aforementioned algorithm 
we have performed the following simple calculation. An inverse Gaussian potential
\[
V(x)= e^\frac{-x^2}{2}
\]
 with half width of $2 \sqrt{(2ln(2))}$
has been constructed which does not support any excited bound states.
The Lanczos algorithm has been used to determine the  eigenstates
of  the corresponding Hamiltonian operator in one dimension using
\[
\phi_1(x)=\langle x|1\rangle = (\frac{2}{\pi})^{1/4}e^{-x^2}
\]
 as the normalized start vector.   
After 18 iterations   the Lanczos algorithm yielded the ground state at $e_{18\ 1}$ = -0.475917 plus a 
spurious state at $e_{18\ 2}$  = 0.529612. In the case of the ground state $\Delta_{18\ 1} =0.0218906$ while
$\Delta_{18\ 2}=2.09673$ clearly indicating that the excited state is spurious.    
The Waxman algorithm, using $u_{1}(x) = 1$ for the determination of the the interpolating function $\epsilon(\lambda)$, 
yielded the ground state at energy = -0.479203. 
Here the aforementioned iterations were repeated until  $\lambda = 1$ yielded  a value to within $10^{-3}$. 
When the Waxman algorithm was used to find the first  excited state it did not yield
 a solution  for $\lambda= 1$. Hence no spurious solutions were obtained with the algorithm. 
 Only for $\lambda\geq 1.35348$ did the Gaussian potential support at least one excited bound state .
   
The spurious states arise in the case of the Lanczos algorithm because  the resulting matrix 
representation of the Hamiltonian operator in the Lanczos basis corresponds to projecting it onto the 
space $\mathcal{F}$. The  diagonalization of  the resulting projected operator yields spurious eigenpairs. 
In the Waxman algorithm, iterations in each step are performed with the Hamiltonian 
operator and no projection or diagonalization is required. Hence, ultimately only the exact bound states 
are obtained and there are no problems with spurious states.

\end{document}